\documentstyle[aps,twocolumn,prl]{revtex}
\begin{document}
\draft
\title {Playing a quantum game with a corrupted source} 
\author{Neil F. Johnson}  
\address {Physics Department and Center for Quantum Computation, 
\\ Clarendon Laboratory, Oxford University, Parks Road, Oxford, OX1 3PU,
U.K.}
%
\date{\today}
\maketitle


\begin{abstract} The quantum advantage arising in a simplified multi-player
quantum game, is found to be a {\em disadvantage} when the game's
qubit-source is corrupted by a noisy `demon'. Above a critical value of  the
corruption-rate, or noise-level, the coherent quantum effects impede the
players  to such an extent that the `optimal' choice of game changes from
quantum to classical.

\end{abstract}

\centerline{\em To appear in Phys. Rev. A (Rapid Comm.)}

\vskip\baselineskip


Information plays a fundamental role in both quantum mechanics\cite{artur}
and games\cite{binmore}. Recently, some pioneering advances have been
made in the field of quantum games\cite{eisert,meyer}.  Eisert et al.
\cite{eisert} considered a quantum version of the famous
$N=2$ player Prisoner's Dilemma\cite{binmore}. The game showed a
fascinating `quantum advantage' as a result of a novel payoff equilibrium.
Benjamin and Hayden subsequently argued that this equilibrium\cite{eisert}
results from an asymmetric restriction in the strategy set; with unrestricted
strategies, it is impossible for such special `coherent quantum equilibria'
(CQE) to arise in the maximally entangled
$N=2$ player game\cite{simon1}.   Following our conjecture \cite{private} that
CQE's arise for
$N\geq 3$ players, Benjamin and Hayden \cite{simon2} created a Prisoner's
Dilemma-like game for $N=3$ with a high payoff CQE\cite{simon2}.  This
effect of `two's company, three's a crowd'  is quite familiar in physical systems
(both classical and quantum) where complex behaviors tend to emerge only
for
$N\geq 3$ interacting particles.

In this paper the quantum advantage arising in a simplified multi-player
quantum game, is found to be a {\em disadvantage} when the game's
qubit-source is corrupted by a noisy `demon' whose activity is unknown to the
players.  Above a critical value of  the corruption-rate, or noise-level, the
coherent quantum effects impede the players  to such an extent that the
classical game outperforms the quantum game; given the choice, the
multi-player system  does better if it adopts classical rather than quantum
behavior. 

Following Ref. \cite{simon2}, $N=3$ players (or `agents') each
receive a qubit in state $|0\rangle$ (or 0). The quantum-game qubits pass
through an entangling $\hat J$-gate\cite{eisert,simon2} (see Fig. 1(a)).
Without loss of generality\cite{simon2} we take ${\hat J}=\frac{1}{\sqrt 2}({\hat
I}^{\otimes 3} + i {\hat F}^{\otimes 3})$ where ${\hat F}={\hat
\sigma}_x$. Hence the input state
$|0\rangle\otimes |0\rangle\otimes |0\rangle\equiv |000\rangle$ becomes 
$\frac{1}{\sqrt 2}(|000\rangle + i |111\rangle)$.  The $i$'th player's strategy
$s_i$ is her procedure for deciding which action to play. The strategy profile
$s=(s_1,s_2,s_3)$ assigns one strategy to each player, and an equilibrium is
a strategy profile with a degree of stability\cite{simon2}, e.g. in a Nash
equilibrium no player can improve her expected payoff by unilaterally
changing her strategy. The payoff table (see Fig. 1(b))  bears some
resemblance to the `El Farol' bar-problem\cite{Arthur} -  the analogy is
not strictly correct, however it aids in understanding
the pay-off table.  A (small) bar has seating capacity for
$2$ people, yet three people want to go. Action 0 (1) means don't go (go). 
State
$|000\rangle$ means everyone stayed away. Noone gains, but  noone is
annoyed that others gained while they lost: the net payoff is zero per player.
State
$|100\rangle$ means one person attended, had plenty of seats (i.e. two) but
no company; her payoff is 1. The other two are annoyed that they didn't
attend and gain from the available seat, hence each gets -9. State
$|110\rangle$  means two attend; they each have a seat and have company
so they get 9. The third person, while not getting maximum enjoyment, is at
least relieved that she didn't make the effort to attend (making the bar
overfull); she gets 1. State
$|111\rangle$  means they all attend. They benefit from lots of company but
not enough seating; they all get 2.   Input qubits must be supplied for each
turn of the game.  The players assume that the input qubits are always
$|0\rangle$ (or 0) hence yielding the payoffs in Fig. 1(b).  The classical game
involves only one of two possible states for each player's qubit at each stage
(0 or 1) and hence one particular outcome in the pay-off table of Fig. 1(b). In
the quantum game a superposition of qubit states is possible and hence a
superposition of outcome states will generally arise - the classical game is
therefore embedded in the quantum game.  As in conventional game
theory\cite{binmore}, average payoffs are given by an expectation value over
the possible measurement results. 

Classical game players either leave the input qubit 0 unchanged, or flip it
to 1. Allowing full knowledge of the payoff table, classical game players will
search for the dominant strategy payoff (2,2,2) and hence choose action $1$
\cite{simon2}.  Following the approach to $N=3$ player  classical games
of Ref. \cite{usprl}, each player is assigned a $p$ value where $p$ is the
probability of leaving the input qubit unflipped, i.e. {\em not} flipping the input
qubit. For simplicity, suppose $p=0$, $1/2$ or $1$ instead of being
continuous\cite{usprl}. There are
$3^3=27$ possible profiles or `configurations' $(p_1,p_2,p_3)$. These yield
ten `classes' each containing $C\geq 1$ configurations which are equivalent
under interchange of player label\cite{usprl}. Table I shows the average
payoffs for each configuration class.  Given that the input is 0, the dominant
strategy equilibrium corresponds to all players choosing
$p=0$, i.e. class (iv) in Table I. Hence although the continuous-parameter
$p$-space has been discretized to only three values, this description includes
the desired dominant strategy equilibrium. 
 The quantum game players, having followed the analysis of Ref.
\cite{simon2} in which the special (5,9,5) `quantum' payoff is presented, 
independently decide to play for the CQE given there. In particular, Ref.
\cite{simon2} shows that the strategies $\hat I$,
$\hat\sigma_x$, and 
$\frac{1}{\sqrt 2}(\hat\sigma_x + \hat\sigma_z)$ yield a novel, high payoff
CQE\cite{simon2} given input qubit $|0\rangle$. We will assume that the set
of 
$3^3=27$ strategy profiles formed from these three simple strategies contain
the {\em only}  strategy profiles subsequently chosen by the quantum game
players. Again this choice is restricted - in particular the quantum game
should include {\em all} SU(2) operations\cite{simon1}. However it allows
for a straightforward comparison between quantum and classical games
without the complication of continuous-parameter sets. The resulting Table II
provides a simple quantum analog of Table I.  $\hat\sigma_x$ corresponds to
not qubit-flipping with probability $p=0$, hence we denote it as `$\hat p\equiv
0$'.  $\frac{1}{\sqrt 2}(\hat\sigma_x +
\hat\sigma_z)$ corresponds to  not qubit-flipping with probability $p=1/2$,
hence we denote it as `$\hat p\equiv 1/2$'. 
$\hat I$ corresponds to not qubit-flipping with probability $p=1$, hence we
denote it as `$\hat p\equiv 1$'. (This correspondence can be established by
imagining switching off the $J$-gates). In both quantum and classical 
games, players are unable to communicate between themselves hence
they cannot coordinate which player picks which strategy. In the quantum
game, this is more critical since the CQE (i.e. the Nash equilibrium given by
class (viii) in Table II) involves players using different
$\hat p$'s. (Although class (vii) has the same average payoff
$\langle\hat\$\rangle=19/3$, it is not `fair' to all players and is not a Nash
equilibrium).   The payoffs in Tables I and II  (indicated by $(\dots)$) are in
general quite different, i.e. the quantum and classical systems behave
differently. 

Now consider the effect of a noisy source created by an external
`demon' (Fig. 1(a)). The demon controls the input qubit corruption-level,
however the players are unaware of his presence. This is reminiscent of a
`Crooked House' in gambling -  players assume the source (e.g. deck of
cards) is clean even though it may have been corrupted by the supplier (e.g.
dealer). In Table I (II), the average payoffs with input qubits always 1
($|1\rangle$) are shown as
$[\dots]$. Again, the quantum and classical payoffs are generally quite
different.  Comparing  the
$(\dots)$ and $[\dots]$ entries in column
${\langle\hat\$\rangle}$ of Table II,  and repeating this for  column $\$$ of
Table I, we see that the quantum game exhibits a {\em lower} symmetry than
the classical game under interchange of input qubit, e.g.  there are two
entries ${\langle\hat\$\rangle}=(19/3)$ in Table II but only one entry
$[19/3]$.  A remarkable result is obtained if we now assume that the source
contains  equal numbers of  $|0\rangle$ and $|1\rangle$ (or 0 and 1) qubits
on average: the quantum and classical games now produce identical payoffs
for a given class (i.e. $\overline{\langle\hat\$\rangle}=\overline\$$). Also, the
resulting payoff entries for {\em each} $p$-value within a given class become
identical. In short, the quantum and classical games converge to produce
{\em identical} payoffs for a given strategy class.
 
Since the players are unaware of the demon's presence, they will still try to
achieve the dominant strategy equilibrium payoff (2,2,2) for the classical
game, i.e. class (iv) in Table I,  and the superior CQE  payoff (5,9,5) for the
quantum game, i.e. class (viii) in Table II. We now examine the average
payoff from these two classes  to see which game is `optimal' from the
players' collective perspective. Let
$x$ be the input qubit noise-level provided by the demon's supply,
representing the fraction of
$|1\rangle$ (or 1) qubits received by each agent over many turns of the
game. For simplicity we assume that the demon supplies identical qubits at
each turn, i.e. 
$|0\rangle\otimes |0\rangle\otimes |0\rangle$ with probability $(1-x)$ and 
$|1\rangle\otimes |1\rangle\otimes |1\rangle$ with probability $x$. There is no
notion of `memory' so far in the system, hence a periodic qubit sequence 
$\dots |0\rangle |1\rangle |0\rangle |1\rangle |0\rangle |1\rangle$ supplied to
each agent has the same `noise-level' ($x=0.5$) as a random sequence
produced by a memoryless coin-toss.  Class (iv) in Table I yields the average
payoff per player 
$\overline\$_x=0.x+2.(1-x)=2-2x$, while class (viii) in Table II yields
${\overline{\langle\hat\$\rangle}}_x=(-17/3).x+(19/3).(1-x)=19/3 - 12x$.  Figure
2 shows  these average payoffs as a function of  $x$.  There is a crossover at
$x_{\rm cr}= 13/30=0.433$; the quantum game does better than the classical
game for $0\leq x < x_{\rm cr}$ while the classical  game does better than the
quantum game for $x_{\rm cr}< x 
\leq 1$.  If ${\overline{\langle\hat\$\rangle}}_x>2$, and hence $0\leq x < x_-$
where
$x_-=13/36=0.361$, then the quantum game does better than the classical
game even if the demon reduces the classical game noise level to $x=0$.  If
${\overline{\langle\hat\$\rangle}}_x<0$, and hence $x_+< x 
\leq 1$ where
$x_+=19/36=0.528$, then  the classical game will do better than the quantum
game even if the demon increases the classical game noise level to $x=1$. 
Suppose the demon is replaced by a heat bath at temperature $T$; using 
the Boltzmann weighting for a two level system (energy separation $\Delta
E$) yields $k_B T=\Delta E ({\rm ln} [(1-x)x^{-1}])^{-1}$. Hence  $k_B T_ {\rm
cr}=3.7\Delta E$, $k_B T_ -=1.75\Delta E$ while $k_B T_ +$ is unobtainable
(i.e. negative). Given the choice, the `optimal' game for the players to play
therefore changes from being quantum to classical as $T$ (i.e. $x$)
increases. For $T> 3.7\Delta E $,  the classical game `takes over' which is
consistent with a simple-minded notion of a crossover from quantum
$\rightarrow$ classical behavior.  From the viewpoint of risk, the class (viii)
quantum-game players have high potential gains but large potential losses -
this can lead to large fluctuations in their momentary wealth depending on the
demon's actions, and hence large risk. By contrast the class (iv)
classical-game players have a smaller risk because of the potentially smaller
wealth fluctuations.  We emphasize that the degradation of the `quantum
advantage' discussed here arises {\em without} any decoherence between
the $J$-gates, i.e.    there is full coherence within the three-player subsystem.
Note that the quantum advantage would also disappear (in a different way) if
the quantum correlations {\em between} the $\hat J$ and
$\hat J^\dagger$ gates were destroyed,  but this is a trivial limit.

An interesting generalization is to consider an {\em evolutionary} quantum
game in which players may modify their strategies based on
information from the past, i.e. they `learn' from past mistakes \cite{usprl}.  
This introduces a `memory' into the system and allows transitions between
classes in Tables I and II. The memory in the evolutionary version will have a
non-trivial effect on  whether the quantum game outperforms the classical
one, or vice versa \cite{private}; the quantum and/or classical
game\cite{freezing}  may even {\em freeze} into a given configuration.  A
deeper understanding of the relative `advantage' between such classical and
quantum many-player dynamical games may
eventually shed light on  connections between quantum and classical
many-particle, dynamical systems: it is possible that pay-offs can be used to
represent energies,  the entangled state of the many-player quantum game
can represent some exotic many-particle wavefunction, and the demon's
actions can mimic environmental decoherence.  
Interestingly, Frieden et al.
\cite{frieden} have proposed that physical laws are derived from an extremum
principle for the Fisher information of a measurement and the information
bound in the physical quantity being measured\cite{frieden,binder} - this EPI
(Extreme Physical Information) principle represents a game played
against Nature. Since the observer can never win\cite{frieden}, the
phenomenon of interest takes on an all-powerful, but malevolent, force - this
is the information `demon' who is looking to increase the degree of `blur' of
information, and against whom the players are forced to play.

I am very grateful to Simon Benjamin for his continued collaboration. I also
thank Seth Lloyd, Philippe Binder, Pak Ming Hui and Luis Quiroga for
discussions.


\vskip\baselineskip



\centerline{\bf Figure Captions}

\bigskip

\noindent Figure 1:  Three-player game: a) classical (top) and quantum (bottom) with
input qubits/bits supplied by a (demonic) external source.  b)
Payoff table.

\bigskip

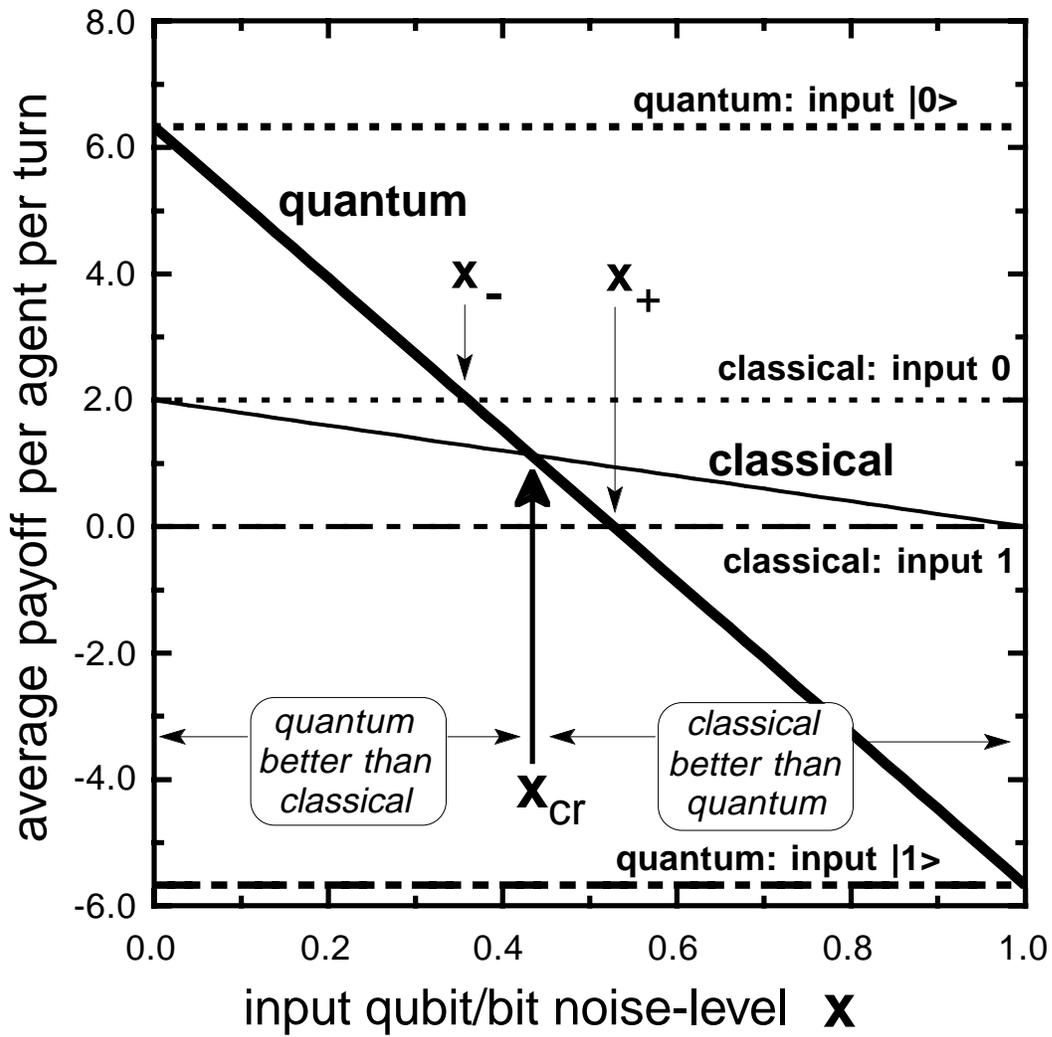
\noindent Figure 2:  Average payoff per player (`agent') per turn for quantum
game (thick solid line) and classical game (thin solid line) as a function of input qubit/bit
noise-level $x$ (i.e. demon's corruption-rate). Dotted lines correspond to payoff for
pure
$|0\rangle$ (or 0) input, while dotted-dashed lines are for pure $|1\rangle$ (or 1) input.

\bigskip

\clearpage

\begin{minipage}{6in}

\begin{table}

\label{table1}
\caption{Average payoffs for classical game (players or `agents' denoted by `a'). 
$p$ is probability of {\em not} flipping the input qubit.  Average payoffs for
input qubit 0 are shown as
$(\dots)$, those for input qubit 1 are shown as $[\dots]$, while those for $50:50$
mixture (i.e. $x=0.5$) of input qubits are shown without parentheses.
$\$$ is
payoff averaged over the three players, for a given input qubit (0 or 1).  
$\overline\$$ is $\$$ averaged over input qubit.}
\bigskip

\begin{tabular}{ccccccc} Class & $p=0$ & $p=1/2$ & $p=1$ & $C$ & 
$\$$ &  $\overline\$$
\\
\tableline  
i) & \_ & aaa(1/2)[1/2]1/2 & \_ & 1 & (1/2)[1/2] & 1/2 \\ 
ii) & a(21/4)[-17/4]1/2 & aa(3/4)[1/4]1/2 & \_ & 3 & (9/4)[-5/4] & 1/2 \\  
iii) & aa(11/2)[-9/2]1/2 & a(3/2)[1/2]1 & \_ & 3 & (25/6)[-17/6] & 2/3 \\  
iv) & aaa(2)[0]1 & \_ & \_ & 1 & (2)[0] & 1 \\  
v) & \_ & \_ & aaa(0)[2]1 & 1 & (0)[2] & 1\\  
vi) & a(1)[1]1 & \_ & aa(-9)[9]0 & 3 & (-17/3)[19/3] & 1/3 \\
vii) & aa(9)[-9]0 & \_ & a(1)[1]1 & 3 & (19/3)[-17/3] & 1/3 \\  
viii)& a(5)[-4]1/2 & a(0)[0]0& a(-4)[5]1/2& 6 & (1/3)[1/3] & 1/3\\  
ix)& \_ & aa(1/4)[3/4]1/2 & a(-17/4)[21/4]1/2 & 3 & (-5/4)[9/4]  & 1/2\\  
x) & \_ & a(1/2)[3/2]1 & aa(-9/2)[11/2]1/2 & 3 & (-17/6)[25/6] & 2/3
\\
\end{tabular}

\end{table}

\vskip\baselineskip

\begin{table}

\label{table2}
\caption{Average payoffs for quantum game (players or `agents' denoted by `a'). 
$\hat p\equiv 0$ corresponds to 
$\hat\sigma_x$; $\hat p\equiv1/2$ corresponds to $1/{\sqrt
2}(\hat\sigma_x+\hat\sigma_z)$;
$\hat p\equiv 1$ corresponds to
$\hat I$ (see text). Average payoffs for
input qubit $|0\rangle$ are shown as
$(\dots)$, those for input qubit $|1\rangle$ are shown as $[\dots]$, while those for
$50:50$ mixture (i.e. $x=0.5$) of input qubits are shown without parentheses.
$\langle\hat\$\rangle$ is
payoff averaged over the three players for a given input qubit ($|0\rangle$ or
$|1\rangle$).  
$\overline{\langle\hat\$\rangle}$ is $\langle\hat\$\rangle$ averaged over input qubit.}

\bigskip

\begin{tabular}{ccccccc} Class & ${\hat p}\equiv 0$ & ${\hat p}\equiv1/2$   & ${\hat
p}\equiv 1$ &
$C$ & 
$\langle\hat\$\rangle$ &  $\overline{\langle\hat\$\rangle}$ 
\\
\tableline  
i) & \_ & aaa(-15/4)[19/4]1/2 & \_ & 1 & (-15/4)[19/4] & 1/2 \\ 
ii) & a(-15/4)[19/4]1/2 & aa(-15/4)[19/4]1/2 & \_ & 3 & (-15/4)[19/4] & 1/2 \\  
iii) & aa(-7/2)[9/2]1/2 & a(3/2)[1/2]1 & \_ & 3 & (-11/6)[19/6] & 2/3 \\  
iv) & aaa(2)[0]1 & \_ & \_ & 1 & (2)[0] & 1 \\  
v) & \_ & \_ & aaa(0)[2]1 & 1 & (0)[2] & 1\\  
vi) & a(1)[1]1 & \_ & aa(-9)[9]0 & 3 & (-17/3)[19/3] & 1/3 \\
vii) & aa(9)[-9]0 & \_ & a(1)[1]1 & 3 & (19/3)[-17/3] & 1/3 \\  
viii)& a(5)[-4]1/2 & a(9)[-9]0 & a(5)[-4]1/2& 6 & (19/3)[-17/3] & 1/3\\  
ix)& \_ & aa(19/4)[-15/4]1/2 & a(19/4)[-15/4]1/2 & 3 & (19/4)[-15/4]  & 1/2\\  
x) & \_ & a(3/2)[1/2]1 & aa(-7/2)[9/2]1/2 & 3 & (-11/6)[19/6] & 2/3
\\
\end{tabular}

\end{table}

\end{minipage}

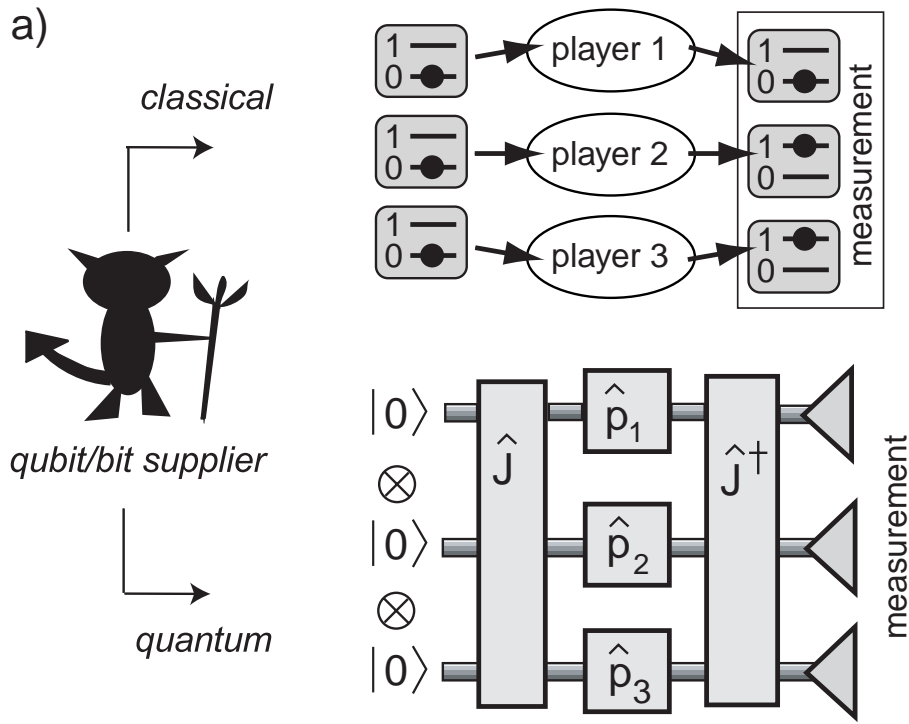
\end{document}